\documentclass[a4paper,11pt]{article}
\pdfoutput=1

\usepackage{jinstpub}
\usepackage{lineno}
\usepackage{subcaption}

\proceeding{Light Detection In Noble Elements - LIDINE 2025\\
21–24 October 2025\\
Hong Kong, China}

\title{\boldmath Geometry-driven impact of photosensor placement on S2-based XY reconstruction in a dual-phase argon TPC}

\author[a,b]{Jilong Yin}
\author[a,b,1]{and Yi Wang\note{Corresponding author.}}
\affiliation[a]{Institute of High Energy Physics, Chinese Academy of Sciences,\\Beijing 100049, China}
\affiliation[b]{University of Chinese Academy of Sciences,\\Beijing 100049, China}

\emailAdd{wangyi90@ihep.ac.cn}

\abstract{
Accurate reconstruction of the horizontal vertex $(x,y)$ from the S2 electroluminescence pattern is essential for fiducialization and background rejection in dual-phase argon time projection chambers. In this work, we perform a Geant4-based simulation study using the \texttt{G4DS} framework to investigate how detector geometry, in particular the distance between the top photodetector plane and the gas pocket, impacts S2-based XY reconstruction. A compact dual-phase argon TPC instrumented with seven Hamamatsu R8520-506 PMTs is simulated with electron recoils at 41.5~keV (corresponding to the ${}^{83m}\mathrm{Kr}$ calibration energy), as well as 1.0~keV to probe the low-S2 regime. The PMT array height is scanned from 0~mm to 50~mm, and XY positions are reconstructed using a geometrical solid-angle (GSA) method with the S2 emission modeled by 1~mm-thick slices across the 7~mm gas pocket. The results show a clear non-monotonic dependence of reconstruction bias and resolution on PMT height, driven by the trade-off between S2 light sharing and photon statistics. These findings provide guidance for geometry optimization in future low-threshold dual-phase argon detectors and will be validated with upcoming prototype measurements.
}


\begin{document}
\maketitle
\flushbottom

\section{Introduction and motivation}
\label{sec:intro}

Dual-phase argon time projection chambers (TPCs) rely on the spatial distribution of the electroluminescence (S2) signal to reconstruct the horizontal $(x,y)$ position of interaction vertices. Accurate XY reconstruction is essential for fiducialization and background rejection, and is therefore a key performance parameter for dark matter searches using dual-phase TPCs~\cite{ref1,ref2,ref3}.

In practice, according to the results from major xenon- and argon-based experiments~\cite{ref4,ref5}, the quality of S2-based XY reconstruction depends not only on photoelectron statistics and reconstruction algorithms, but also on the geometrical configuration of the detector~\cite{ref6}. In particular, the relative placement of the top photodetector array with respect to the S2 emission region influences how the S2 light is shared among channels. If the photodetectors are positioned too close to the gas pocket, the S2 signal can become highly concentrated in one channel, reducing the sensitivity of the observed light pattern to lateral position changes. Conversely, placing the photodetectors too far away reduces the total collected light, which can also degrade reconstruction performance.

These effects are expected to be especially relevant at low recoil energies, where the S2 signal size is intrinsically small and robust XY reconstruction becomes more challenging~\cite{ref7}. Motivated by these considerations, this work investigates how detector geometry---specifically, the distance between the photodetector plane and the gas pocket---affects the performance of S2-based XY reconstruction in a dual-phase argon TPC. Using Geant4 simulations, we apply a geometrical solid-angle method over a range of detector configurations and deposited energies, with the goal of identifying geometrical conditions that optimize reconstruction accuracy for future low-threshold argon detectors, for instance DarkSide-LowMass~\cite{ref8}.

\section{Geant4 simulation setup}
\label{sec:sim}

The simulation of S2 electroluminescence and photon detection in a compact dual-phase argon TPC is performed using the \texttt{G4DS} framework based on Geant4~\cite{ref9}. The goal is to generate event samples with known ground-truth vertex positions and corresponding S2 photoelectron (PE) patterns on the top photosensor array, in order to evaluate the dependence of XY reconstruction on detector geometry.

\subsection{Geometry and optical model}

The modeled TPC consists of a cylindrical liquid argon (LAr) target volume with a diameter of 80~mm and a height of 76.4~mm. Above the liquid surface, the gas electroluminescence region (gas pocket filled with gas argon) is modeled as a cylinder with a diameter of 90~mm and a height of 7~mm. The top of the gas pocket is defined by a 2~mm quartz window whose bottom surface is coated with tetraphenyl butadiene (TPB).
The electric field configuration is set to match the DarkSide-50 detector: 200~V/cm in the drift region and 4.2~kV/cm in the electroluminescence region~\cite{ref4}.

\begin{figure}[htbp]
  \centering
  \includegraphics[width=0.6\textwidth]{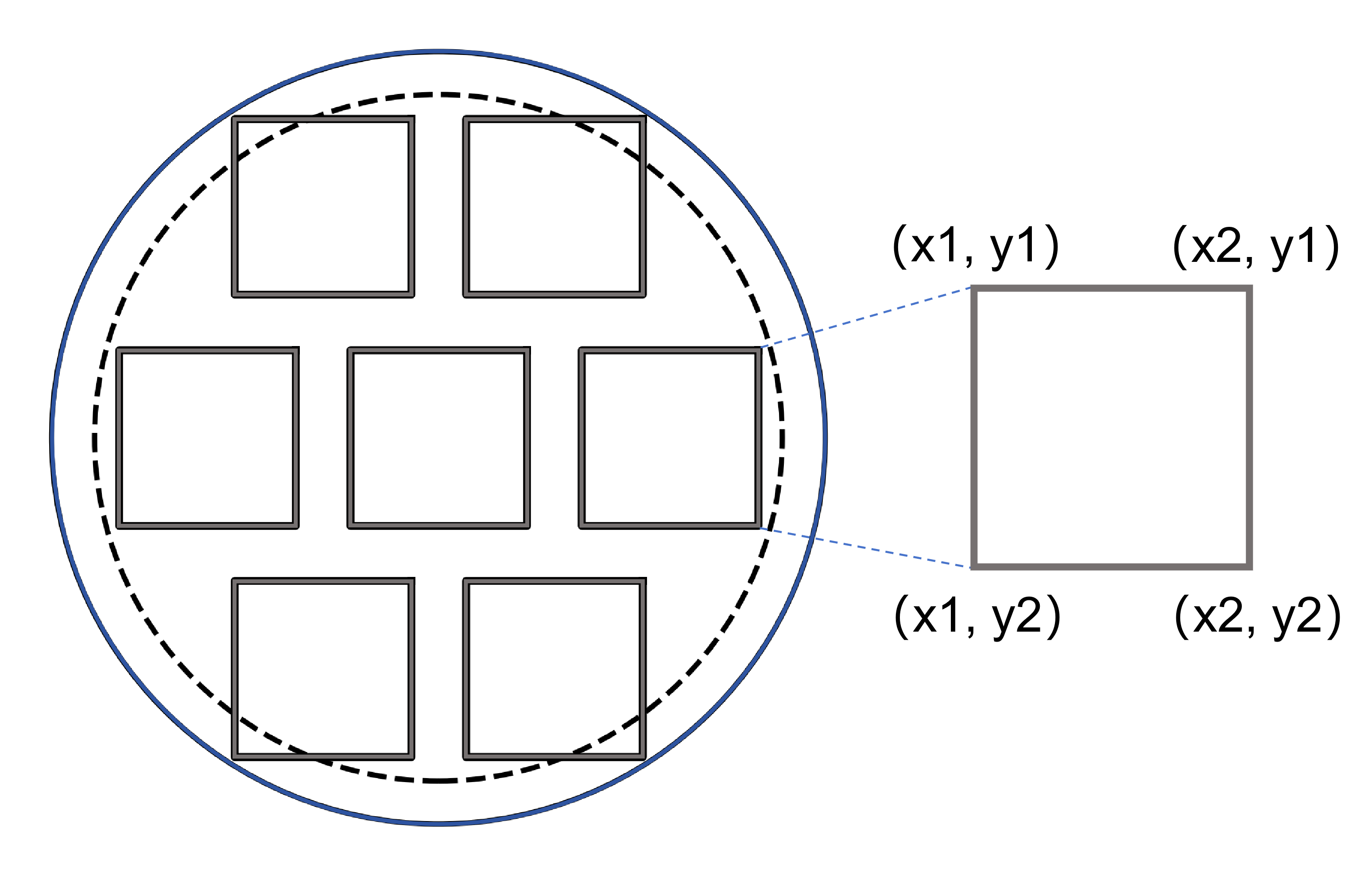}
  \caption{Schematic of the seven PMT array. The square solid line refer to the photocathode area of each PMT; the solid circle refer to the area of the gas pocket; and the dashed circle refer to the area of the liquid argon. The right side zoom-in PMT shows how to define the coordinates of each corner which is used in the algorithm mentioned in section~\ref{subsec:3.1}. }
  \label{fig:1}
\end{figure}

As shown in figure~\ref{fig:1}, seven Hamamatsu R8520-506 photomultiplier tubes (PMTs) are placed above the gas pocket, facing downward, arranged with one central channel and six surrounding channels. In the simulation, each PMT has effective photocathode area with a $20.5~\mathrm{mm}\times 20.5~\mathrm{mm}$ size~\cite{ref10}. The center-to-center spacing between adjacent PMTs is 27~mm.

We define $h$ as the vertical distance between the PMT photocathode plane and a reference plane located 3~mm above the top surface of the gas pocket (i.e. 1~mm above the top surface of the quartz window). The PMT photocathode plane is scanned from $h=0$~mm to 50~mm in steps of 5~mm.

To emulate a reflective inner surface, the inner wall of the LAr volume is covered by enhanced specular reflector (ESR) with TPB coating. In the simulation, this is implemented as an optical surface with reflectivity 0.98 in the relevant wavelength range around 420~nm.

\subsection{Event generation and datasets}

Electron recoil (ER) events are generated uniformly in the XY direction, while the Z coordinate is fixed in the LAr volume. ER energy of 41.5~keV is used in liquid argon, corresponding to the combined ${}^{83m}\mathrm{Kr}$ calibration energy (9.4~keV + 32.1~keV), which is a commonly used internal calibration source for liquid argon detectors~\cite{ref11}. In addition, lower-energy events at 1.0~keV electron recoil are generated to evaluate reconstruction performance in the small-S2 regime relevant for low-energy recoils.

The simulation includes energy deposition, ionization electron generation, and the subsequent spread of the charge cloud due to diffusion during transport in LAr (the diffusion difference caused by varying drift length is not taken into account since all events are generated at a fixed Z) and in the gas region~\cite{ref12}. For each configuration, the detected S2 response in each PMT channel is recorded to construct the PE pattern used as input to the reconstruction algorithms. The quantum efficiency of the PMTs is set to 26\% as suggested by the datasheet~\cite{ref10}.

For each $h$ configuration, 100,000 events are generated. For every event, the ground-truth position $(x_{\mathrm{true}},y_{\mathrm{true}})$ and the corresponding PMT response $\{n_i\}$ are stored, enabling a direct evaluation of reconstruction bias and resolution as functions of height and deposited energy.

\section{XY reconstruction method and results}
\label{sec:reco_results}

In this study, the interaction vertex $(x,y)$ is reconstructed using the S2 light-sharing pattern recorded by the seven-channel top PMT array. For each simulated event, the reconstruction algorithm takes the channel responses $\{n_i\}$ as input and returns $(x_{\mathrm{rec}},y_{\mathrm{rec}})$, which is compared to the ground-truth position $(x_{\mathrm{true}},y_{\mathrm{true}})$ from simulation.

We define the reconstruction residuals as
\begin{equation}
\Delta x = x_{\mathrm{rec}} - x_{\mathrm{true}}, \quad 
\Delta y = y_{\mathrm{rec}} - y_{\mathrm{true}},
\end{equation}
and the radial deviation
\begin{equation}
\Delta r = \sqrt{\Delta x^{2}+\Delta y^{2}} .
\end{equation}
The reconstruction bias is evaluated with the mean of $\Delta r$, while the reconstruction resolution (precision) is characterized by the standard deviation of $\Delta r$ within the selected sample.

\subsection{Geometrical solid-angle (GSA) method}
\label{subsec:3.1}

In the GSA approach, the expected S2 light-sharing pattern is approximated using the geometrical solid angle subtended by each PMT channel with respect to an assumed S2 emission location. For a given hypothesis position $(x,y)$, the expected fraction observed by channel $i$ is
\begin{equation}
P_i(x,y) = \frac{\Omega_i(x,y)}{\sum_{j=1}^{N}\Omega_j(x,y)}, \quad N=7,
\end{equation}
where $\Omega_i$ is computed using the analytic expression for a square aperture~\cite{ref13}:
\begin{equation}
\omega(x,y)=\arctan\left(\frac{xy}{z\sqrt{x^2+y^2+z^2}}\right),
\end{equation}
\begin{equation}
\Omega_i =4\left[\omega(x_2,y_2)-\omega(x_1,y_2)-\omega(x_2,y_1)+\omega(x_1,y_1)\right].
\end{equation}
Here $z$ is the distance from the emission point to the photocathode plane, and $(x_1,y_1)$, $(x_1,y_2)$, $(x_2,y_1)$, and $(x_2,y_2)$ denote the four corners of the square photocathode of each PMT, expressed in a coordinate system illustrated in figure~\ref{fig:1}.

Since S2 electroluminescence is produced throughout the gas pocket rather than at a single point, we model the emission region as an extended source along $z$. The gas gap thickness is 7~mm and is divided into $K=7$ uniform slices with thickness $\Delta z = 1$~mm. The expected light-sharing pattern is obtained by summing the solid-angle contributions from the slice midpoints:
\begin{equation}
P_i(x,y)=\frac{\sum_{k=1}^{K} \Omega_i^{(k)}(x,y)}
{\sum_{j=1}^{N}\sum_{k=1}^{K} \Omega_j^{(k)}(x,y)},
\quad N=7, K=7,
\end{equation}
where $\Omega_i^{(k)}$ is computed using the corresponding slice midpoint distance $z_k$. In this work, the electroluminescence yield is assumed to be constant across the gas gap, and therefore all slices are assigned equal weight.

The observed PE pattern is normalized as $\hat{p}_i = n_i/\sum_j n_j$, and the reconstructed position is obtained by minimizing
\begin{equation}
\chi^2(x,y)=\sum_{i=1}^{N}\left(\hat{p}_i-P_i(x,y)\right)^2 .
\end{equation}

\subsection{Results and discussion}

\begin{figure}[htbp]
  \centering

  \begin{subfigure}[b]{0.48\textwidth}
    \centering
    \includegraphics[width=\textwidth]{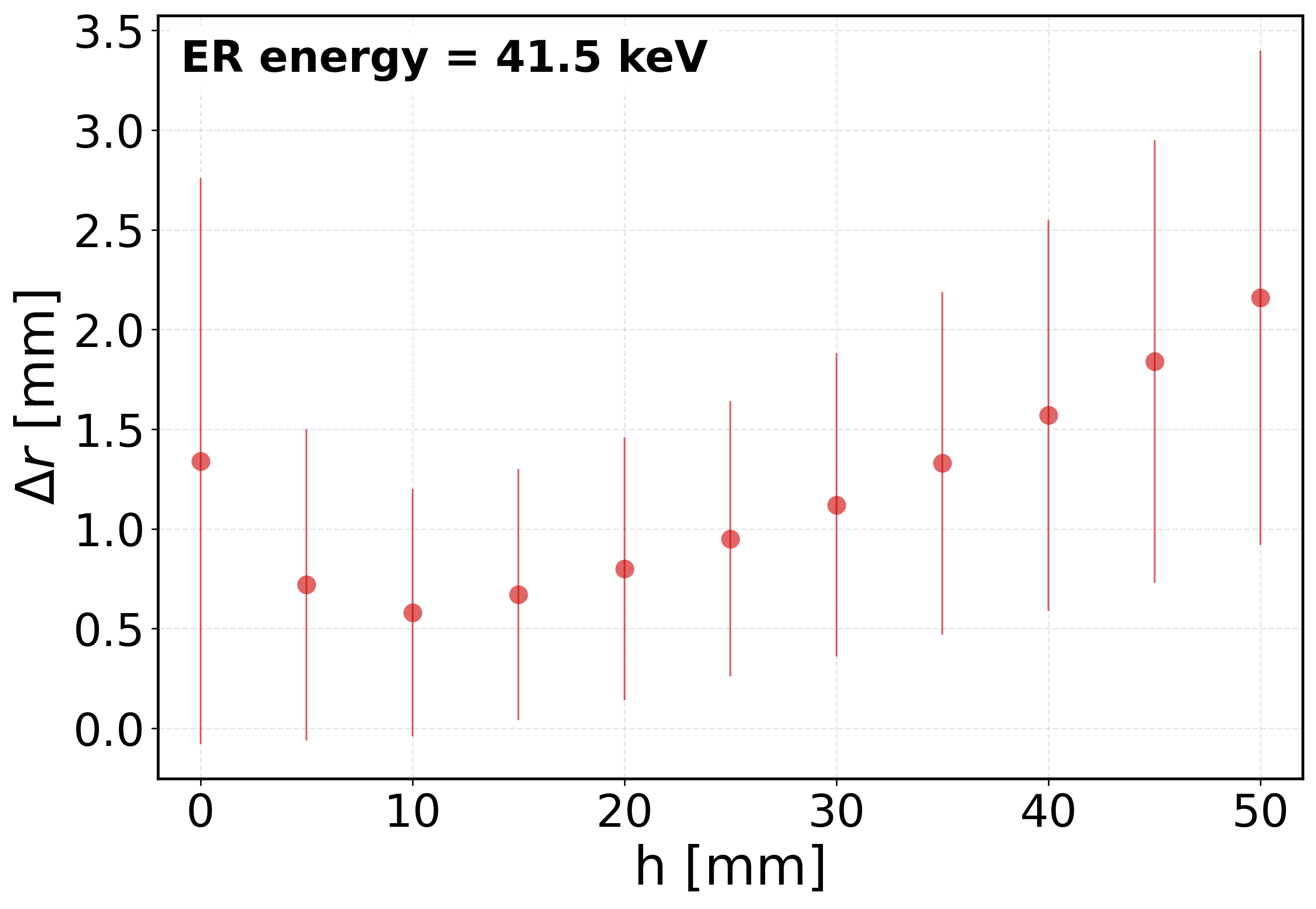}
    \caption{ER energy is 41.5~keV.}
    \label{fig:2a}
  \end{subfigure}
  \hfill
  \begin{subfigure}[b]{0.48\textwidth}
    \centering
    \includegraphics[width=\textwidth]{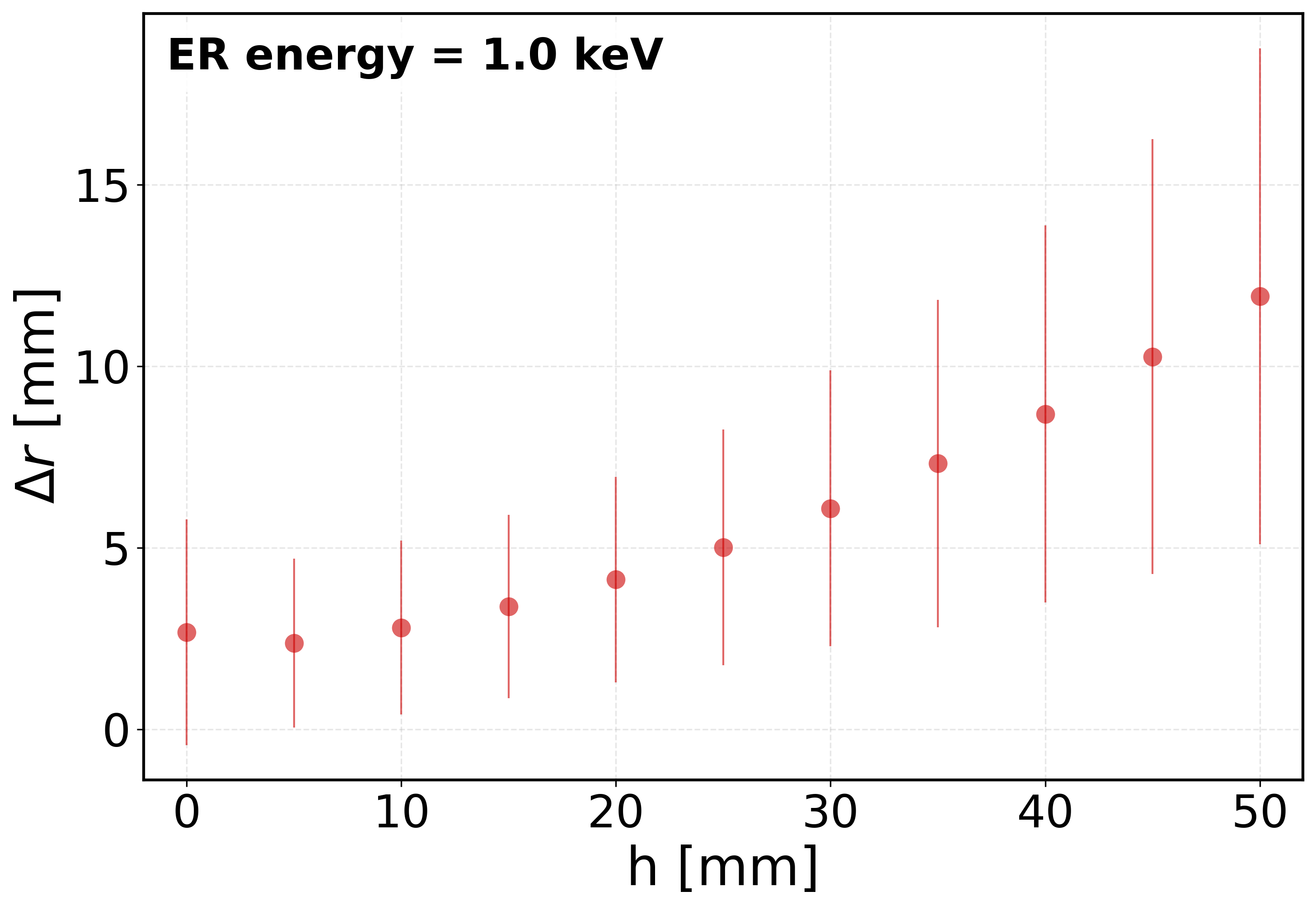}
    \caption{ER energy is 1.0~keV.}
    \label{fig:2b}
  \end{subfigure}

  \caption{Reconstruction bias as a function of the PMT height $h$ for two ER energies. Markers show the mean radial deviation $\langle\Delta r\rangle$ and the error bars indicate one standard deviation of $\Delta r$.}
  \label{fig:2}
\end{figure}

Figure~\ref{fig:2} summarizes the reconstruction performance as a function of $h$ in terms of the mean radial deviation $\langle\Delta r\rangle$ (bias) and the spread of $\Delta r$ (resolution). With different ER energies deposited, we observe a clear non-monotonic dependence: both bias and resolution degrade when the PMTs are placed very close to the gas pocket, improve as $h$ increases, and degrade again when $h$ becomes too large.

This behavior is driven by a competition between light-sharing sensitivity and photon statistics. At small $h$, the detected S2 signal is highly concentrated in the nearest channel, producing weak position dependence in the normalized PE pattern and resulting in degraded reconstruction. At large $h$, the total collected PE decreases and the channel responses become more similar, again reducing the information content available for position inference. For 41.5~keV and 1.0~keV, the optimal $h$ is 10 mm and 5 mm respectively.

\begin{figure}[htbp]
  \centering

  \begin{subfigure}[b]{0.48\textwidth}
    \centering
    \includegraphics[width=\textwidth]{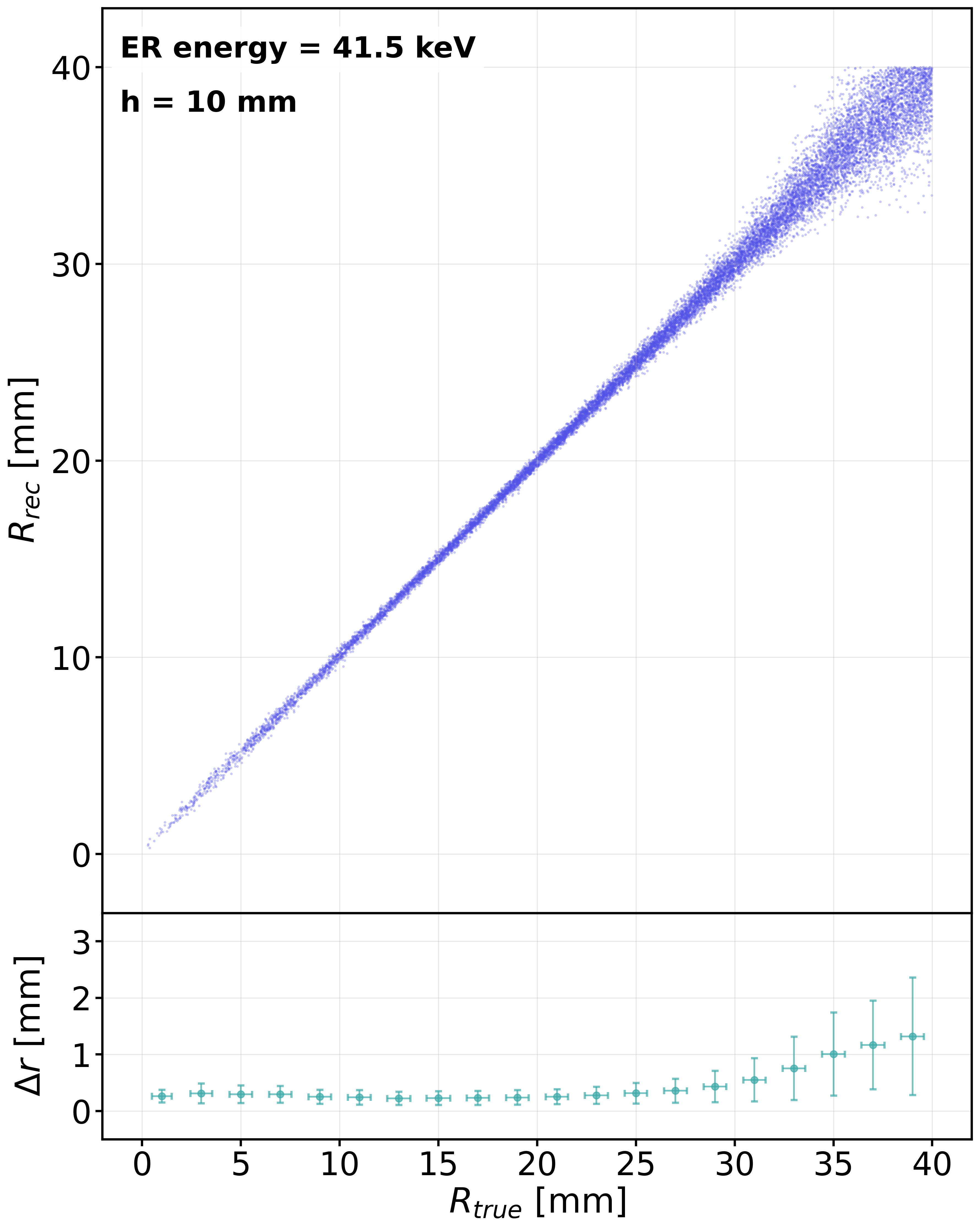}
    \caption{ER energy is 41.5~keV; $h$ is 10~mm.}
    \label{fig:3a}
  \end{subfigure}
  \hfill
  \begin{subfigure}[b]{0.48\textwidth}
    \centering
    \includegraphics[width=\textwidth]{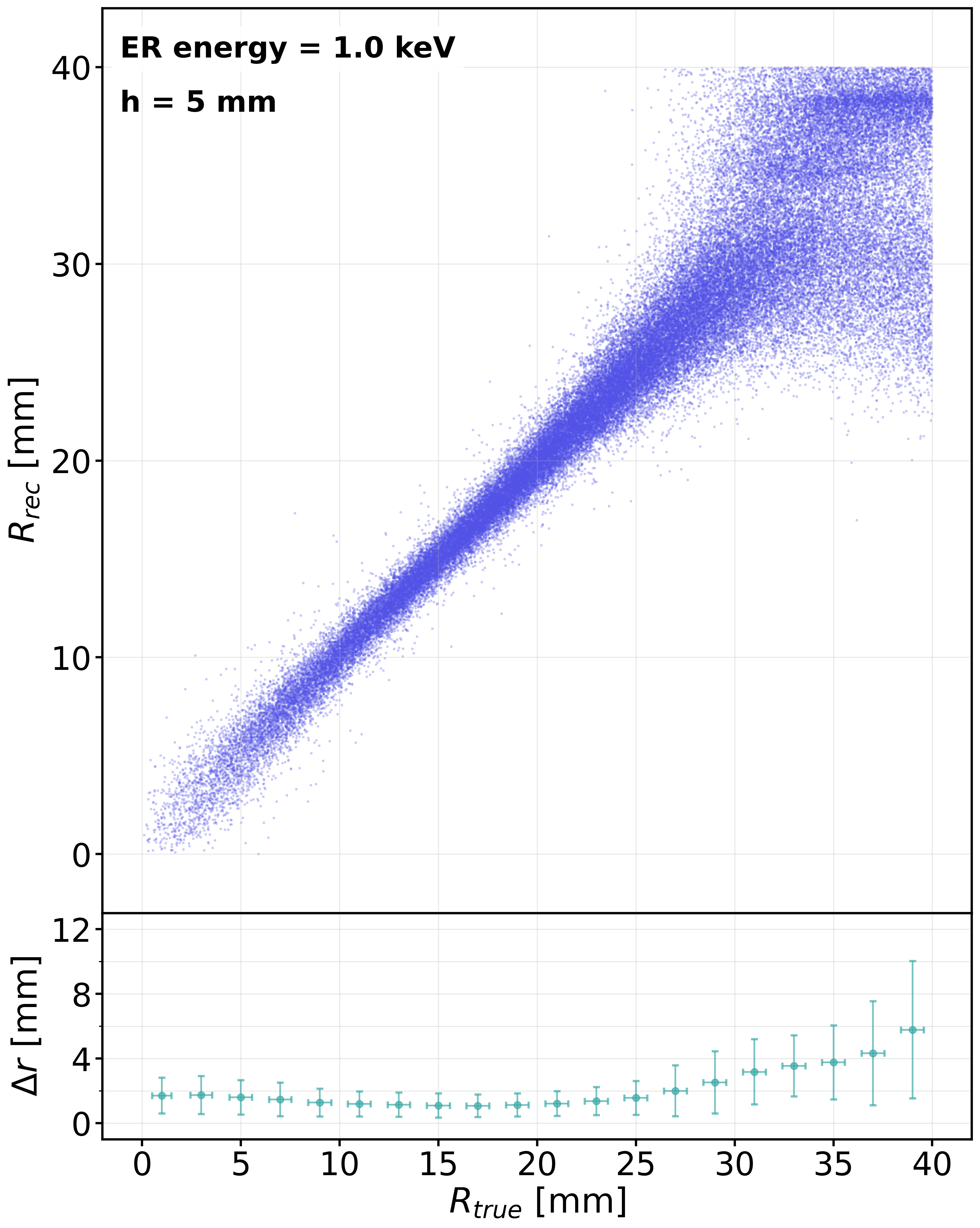}
    \caption{ER energy is 1.0~keV; $h$ is 5~mm.}
    \label{fig:3b}
  \end{subfigure}

  \vspace{0.5em}

  \begin{subfigure}[b]{0.48\textwidth}
    \centering
    \includegraphics[width=\textwidth]{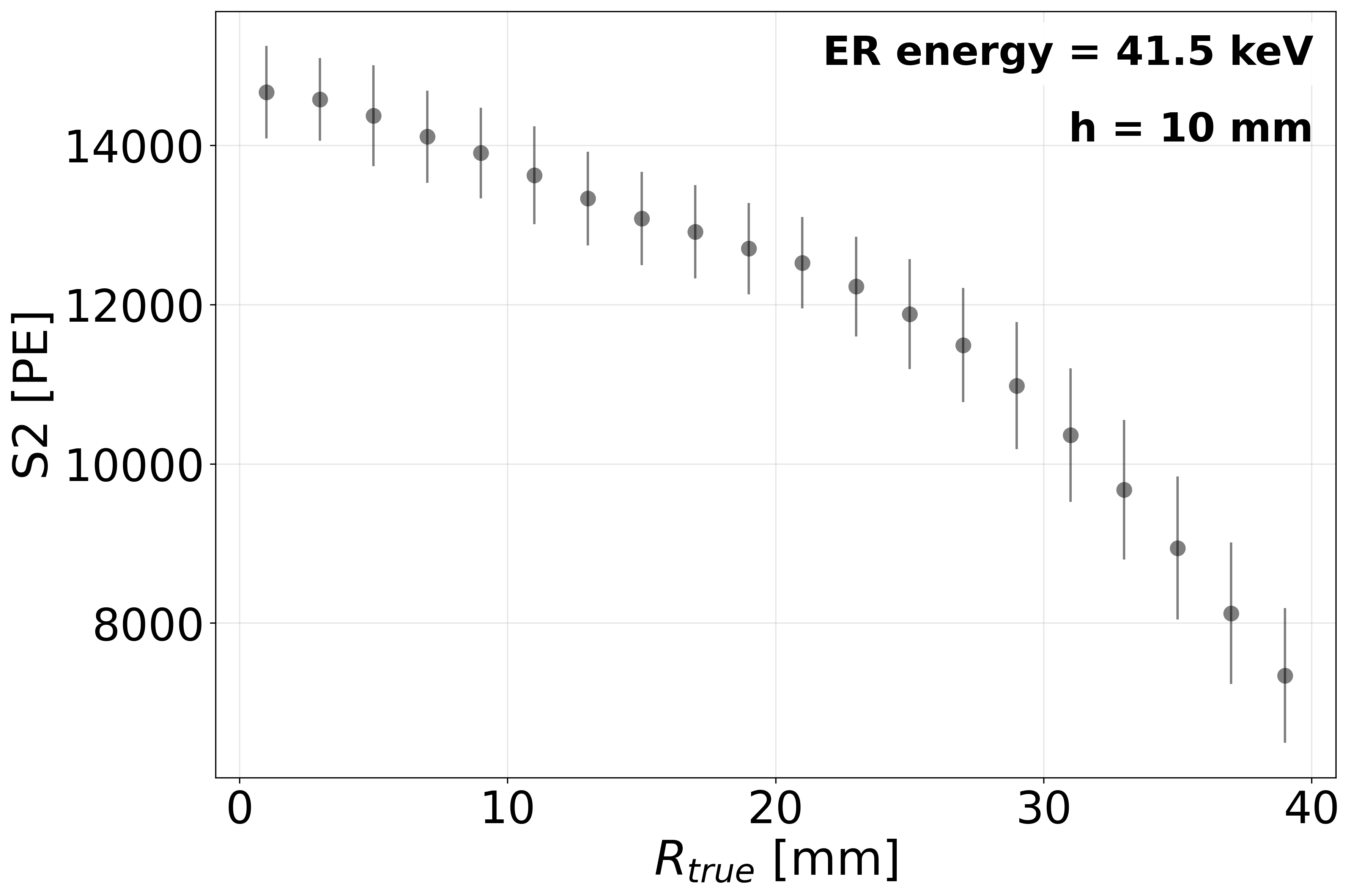}
    \caption{ER energy is 41.5~keV; $h$ is 10~mm.}
    \label{fig:3c}
  \end{subfigure}
  \hfill
  \begin{subfigure}[b]{0.48\textwidth}
    \centering
    \includegraphics[width=\textwidth]{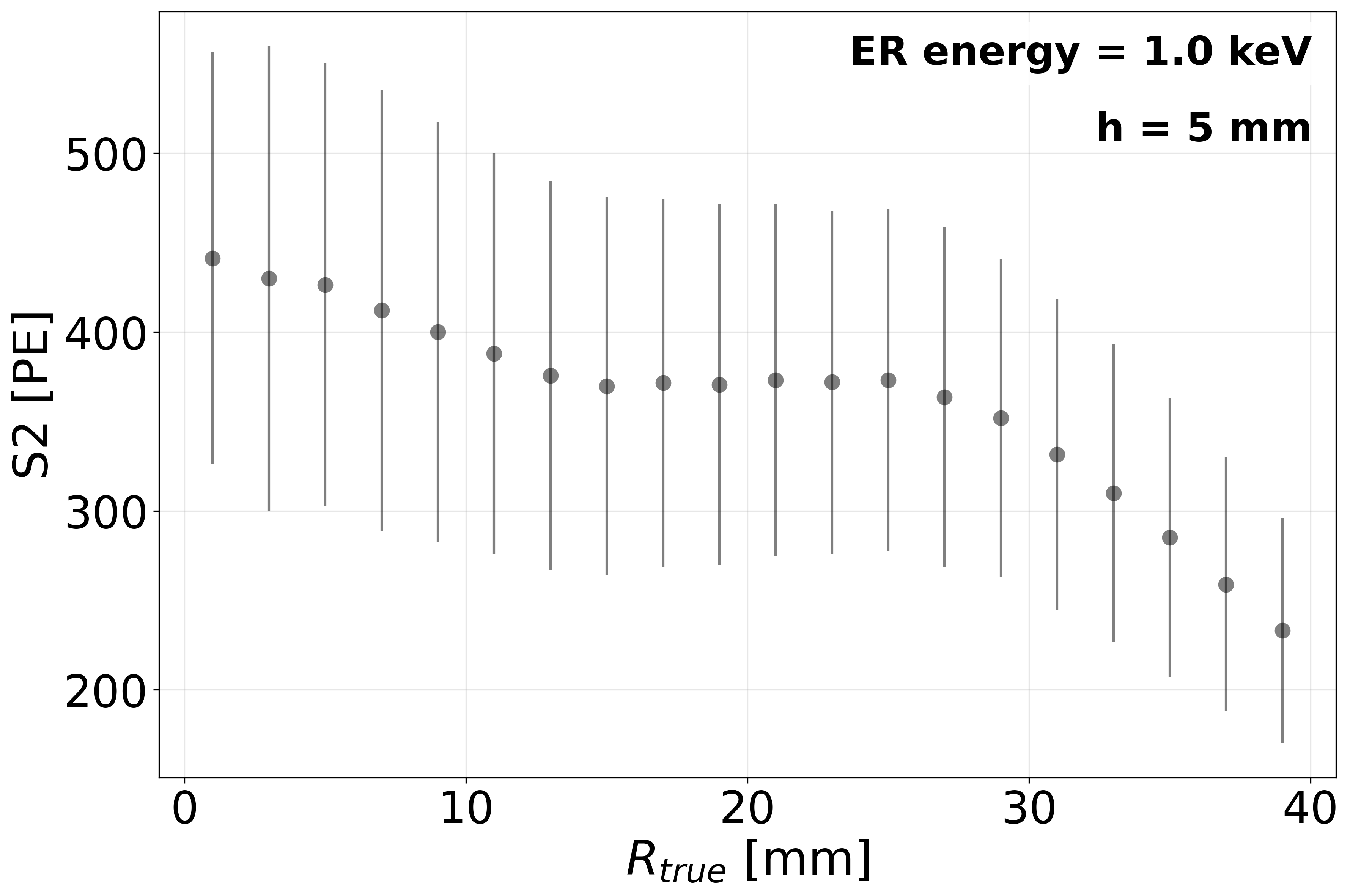}
    \caption{ER energy is 1.0~keV; $h$ is 5~mm.}
    \label{fig:3d}
  \end{subfigure}

  \caption{Reconstruction performance and S2 size versus the true radial position $R$. For each ER energy, $h$ is set to the corresponding near-optimal value inferred from figure~\ref{fig:2}. Figure~\ref{fig:3a} and~\ref{fig:3b} show $R_{\mathrm{rec}}$ versus $R_{\mathrm{true}}$ and the binned mean bias (with error bars) as a function of $R_{\mathrm{true}}$. Figure~\ref{fig:3c} and~\ref{fig:3d} show the total detected S2 photoelectrons as a function of $R_{\mathrm{true}}$.}
  \label{fig:3}
\end{figure}

Figure~\ref{fig:3} illustrates the reconstruction performance versus the true radial position $R$ at the near-optimal heights for the two energies studied. For 41.5~keV, the reconstruction is approximately uniform across most of the active region, while a larger bias is observed near the edge of the liquid volume, where both the S2 size decreases and optical edge effects (e.g. reflections from the reflector) become more important. For 1~keV, the overall performance is worse due to the smaller S2 size, and the degradation near the edge is more pronounced. These results indicate that the geometry-driven optimization can improve the XY reconstruction, but further developments will be needed to reduce the bias and uncertainty in the low-energy regime.

\section{Conclusion and outlook}
\label{sec:conclusion}

In this work, we performed a Geant4-based study of S2-pattern-driven XY reconstruction in a compact dual-phase argon TPC geometry, focusing on the impact of the distance between the photosensors and the gas pocket, using a geometrical solid-angle (GSA) reconstruction algorithm.

The simulation results show a clear non-monotonic dependence of reconstruction performance on photosensor height. When the sensors are too close to the S2 emission region, the S2 light sharing becomes highly non-uniform and the PE pattern is less sensitive to lateral position changes, leading to degraded reconstruction bias and resolution. When the sensors are too far away, reduced photon statistics and increasingly similar channel responses also worsen the reconstruction. As a result, under different recoil energy conditions, an optimal height exists where the trade-off between light-sharing sensitivity and PE statistics yields the best reconstruction precision. In particular, with low recoil energy at 1.0~keV, a few millimeter scale reconstruction is reached and it wll be important for the experiment to explore low recoil energies.

To validate these conclusions, we are preparing a small dual-phase argon TPC prototype with adjustable photosensor height. The GSA method will be tested on experimental data to evaluate its robustness against realistic optical effects, using ${}^{83m}\mathrm{Kr}$ and other low-energy calibration sources such as ${}^{37}\mathrm{Ar}$. In addition, we are developing a calibration technique based on gold-coated cathode spots to provide reference XY positions using time-separated electroluminescence signals (S3), enabling a direct benchmark of reconstruction bias and resolution.

\acknowledgments

This work is supported by the National Key Research and Development Project of China, grant No. 2022YFA1602001.

\end{document}